\documentclass[10pt]{iopart}

\newcommand \bc {\begin{center}}
\newcommand \ec {\end{center}}
\newcommand \ee {\end{equation}}
\newcommand \be {\begin{equation}}
\newcommand \beq {\begin{eqnarray}}
\newcommand \eeq {\end{eqnarray}}
\newcommand \bmu {\begin{multline}}
\newcommand \emu {\end{multline}}
\newcommand \eps {\epsilon}

\begin{document}

\title{Zero-th law in structural glasses: an example} 
\author{Adan Garriga}
\address{Department of Physics, Faculty of Physics, University of
Barcelona\\ Diagonal 647, 08028 Barcelona (Spain). }
\maketitle
\begin{abstract}
We investigate the validity of a zeroth thermodynamic law for
non-equilibrium systems. In order to describe the thermodynamics of
the glassy systems, it has been introduced an extra parameter, the
effective temperature which generalizes the fluctuation-dissipation
theorem (FDT) to off-equilibrium systems and supposedly describes
thermal fluctuations around the aging state. In particular we analyze
two coupled systems of harmonic oscillators with Monte Carlo
dynamics. We study in detail two types of dynamics: sequential
dynamics, where the coupling between the subsystems comes only from
the Hamiltonian; and parallel dynamics where there is another source
of coupling: the dynamics. We show how in the first case the effective
temperatures of the two interacting subsystems are different
asymptotically due to the smallness of the thermal conductivity in the
aging regime. This explains why, in structural glasses, different
interacting degrees of freedom can stay at different effective
temperatures, and never thermalize.
\end{abstract}

\section{INTRODUCTION}

The dynamics of glassy systems has been a subject of intensive
research \cite{SIT,BOU}. In the last decades there has been an
increasing interest in trying to extend the thermodynamics ideas to
systems which are far out of equilibrium. The general
problem is to quantify non-equilibrium fluctuations within a
thermodynamic approach. A salient feature of systems which are in
equilibrium is the fact that the equilibrium fluctuations and the
linear response functions are related by the so-called
fluctuation-dissipation theorem (FDT) \cite{KUB}, which does not hold
for off-equilibrium systems. Several studies of spin-glass mean-field
models have shown that a generalization of the fluctuation-dissipation
theorem is possible through the definition of the
``fluctuation-dissipation ratio'' (FDR)\cite{CUG,FRA}:

\be
X(t,s) = \frac{TG(t,s)}{\frac{\partial C(t,s)}{\partial s}}  (t\geq s)
\label{eqX},\ee

\noindent
which is equal to 1 in equilibrium. It turns out that the behavior of
the quantity $X(t,s)$ is non trivial in the limit $t,s\to\infty$. If
the lowest time $s$ is sent to infinity the quantity $X(t,s)$ becomes
a non-trivial function of the autocorrelation $C(t,s)$. This strong
statement has been proved to hold in the framework of mean-field
spin glasses \cite{CUG,FRA}. Moreover, it has been recently recognized that
the quantity $X$ is generally related to the Parisi order parameter
$P(q)$ which appears in equilibrium studies of spin-glasses providing a
natural link between the static and dynamical properties \cite{FMPP}.

According to relation (\ref{eqX}) the usual fluctuation-dissipation
relation would be recovered if the temperature into the right hand
side of (\ref{eqX}) were $T/X(t,s)$.  This last ratio receives the
name of effective temperature and it has been shown
\cite{KUR,CUKU,EXH} that it has some properties of a macroscopic
temperature. In fact, a proper thermometer coupled to the slow degrees
of freedom can measure it. The question about the convenience of this
temperature to describe the non-equilibrium behavior has been a
subject of controversy in the last years \cite{NIEW2}. While there are
some evidences (not only theoretical but also experimental
\cite{SIT,EXP}) that the violation of FDT gives a good temperature in
the thermodynamic sense, it is unclear what properties of non-equilibrium temperatures are common to the equilibrium ones.

In this letter we want to analyze how effective temperatures equalize
when two systems out of equilibrium are put in contact. In other
words, we want to study if there exist a zeroth law for
non-equilibrium systems. Let us imagine about a vitrified piece of
silica quenched to the room temperature. Because the glass is
off-equilibrium its effective temperature is higher than room
temperature, but when we touch the glass it is not hotter than the
room temperature. Then we can think that some ``fast'' degrees of
freedom are thermalized to the room temperature while other ``slow''
degrees of freedom remain non-thermalized. Why different interacting
degrees of freedom have not reached thermal equilibrium for sufficient
long times? We believe that the answer lies in the fact that the
conductivity in the aging state can be extremely small, and the two
parts of the system never thermalize\cite{art2}. However, we can
achieve a more deep understanding through a detailed analysis of an
illustrative example as a previous stage to offer more simple and
generic considerations. 

The model is a set of harmonic oscillators
evolving by Monte Carlo dynamics introduced in \cite{BPR}. The
importance of this model relies on the fact that at zero temperature it shows typical features of glasses such as aging
in correlation and response functions.  Our interest will be in
considering two coupled sets of harmonic oscillators. Thus, we can see
how the main observables are affected by the coupling, in particular
how the effective temperature evolves for the two sets of interacting
degrees of freedom (i.e for the two different sets of harmonic
oscillators). The interaction may then appear through the Hamiltonian
or through the Monte Carlo dynamics itself. We will discover that the
effective temperature for the two sets of oscillators depends on how
the coupling is done, and we will understand why in vitreous systems
different degrees of freedom may stay at different temperatures
without thermalising at very long times. In this sense the utility of
the extension of the zeroth thermodynamic law to the non-equilibrium
aging state is questioned.

The paper is organized as follows. Section II describes the main aspects
as well as the interest of the model. Section III describes the two
classes of couplings we have considered. Section IV discusses the results and the physical consequences of our work. The last section presents the conclusions.

\section{A SOLVABLE MODEL OF GLASS}

As a simple model of glass we will consider a system of uncoupled
harmonic oscillators evolving with Monte Carlo dynamics introduced
in~\cite{BPR} and recently revisited in \cite{art1,TEO,FEL}. The
Hamiltonian is:
\begin{equation}
H = \frac{1}{2} K \sum_{i=1}^N x_i^2~~~~~.
\end{equation}
The low-temperature Monte Carlo dynamics of this system shows typical
non-equilibrium features of glassy systems like aging in the
correlation and response functions. The interest of this model is that
at low temperatures the low acceptance rate generates entropic
barriers that made the dynamics of the system extremely slow. The
simplicity of this model makes it exactly solvable yielding a lot of
results about the non-equilibrium behavior.

The Monte Carlo dynamics is implemented by small movements: $ x_i
\rightarrow x_i + r_i/\sqrt{N} $ where $ r_i $ are random variables
Gaussian distributed with zero average and variance $ \Delta^2. $ The
move is accepted according to the Metropolis algorithm with
probability $ W(\Delta E) $ which satisfies detailed balance: $
W(\Delta E) = W(-\Delta E) \exp (-\beta \Delta E ), $ where $ \Delta E
$ is the change in the Hamiltonian. The main features of this system
at zero temperature we are interested in are:

\begin{enumerate}

\item{\em Slow decay of the energy.} The evolution equation for the
energy is Markovian. The asymptotic large-time expansion of this
equation shows that the energy decays logarithmically $E(t)\sim
1/\log(t)$. It can also be seen that the acceptance ratio decays
faster $A(t)\sim 1/(t\,\log(t))$.

\item{\em Aging in correlations and responses.} The correlation function
$C(t,s)$ is defined by:
\begin{equation}
 C(t,s) =\frac{1}{N} \sum_{i=1}^N x_i(t)  x_i(s)~~~~~. 
\end{equation}

The response function $ G(t,s)$ is calculated by applying an external
field to the system. Then, the response function is the variation of
the magnetization $M(t)=\frac{1}{N}\sum_{i=1}^N x_i(t)$ of the system when the field is applied:

\begin{equation}
G(t,s) = \left( \frac{\delta M(t)}{\delta h(s)} \right)_{h=0}~~~~~t>s~.
\label{eqM}
\end{equation} 
The asymptotic scaling
behavior for these two quantities is given by,
\be
C(t,s)=C(s,s)\frac{D(s)}{D(t)}~,~~~~~~~
G(t,s)=G(s,s)\frac{D(s)}{D(t)}\Theta(t-s)~,\label{eqG}
\ee
with $D(t)\sim t(\log^2(t))$ \footnote{In the original paper
\cite{BPR} the logarithmic corrections were estimated to be
$\log^{\frac{1}{2}}(t)$. The correct exponent for the logarithmic
corrections was later evaluated by Th. M. Nieuwenhuizen \cite{TEO}.}
and $C(s,s)=\frac{2E(s)}{K}, G(s,s)=\frac{f(s)}{K}$ where $f(t)$ is a
function that contains the whole dynamics and decays like $1/t$.
\cite{BPR}.

\item{\em The effective temperature.}
As said in the introduction, the effective temperature is defined in
terms of the FDR eq.(\ref{eqX}):

\begin{equation}
T_{\rm eff}(t,s)=\frac{T}{X(t,s)}=\frac{\frac{\partial C(t,s)}{\partial s}}{G(t,s)}\label{eq_Teff}~~~~~.
\end{equation}

As we expected, in equilibrium $X(t,s)=1$ and $T_{\rm
eff}=T$. Moreover, $E(s)=T/2$ and the equipartition theorem is
verified. In the non-equilibrium case, the effective temperature only
depends on the lowest time $s$ for all times $s$ and $t$. This is a characteristic feature of this particular model that is believed to hold for structural glass models in the asymptotic limit $ s \rightarrow \infty $.

At zero temperature when slow motion sets in, the system never
reaches the ground state. In this aging regime the
effective temperature verifies, in the limit $s \longrightarrow
\infty:$

\begin{equation}
T_{\rm eff}(s) = 2E(s) + \frac{2}{f(s)} \frac{\partial E(s)}{\partial s} \longrightarrow 2E(s),
\label{eq_Teff3}
\end{equation}

Equation \ref{eq_Teff3} shows how the equipartition theorem can be
extended to the glassy regime.

\end{enumerate}

\section{TWO COUPLED SYSTEMS OF OSCILLATORS}

Now we consider the case in which we couple two systems of harmonic
oscillators. In this case it is possible to compute analytically
how one system affects the other. The Hamiltonian is:

\begin{equation} H = \frac{K_{1}}{2} \sum_{i=1}^N x_i^2 + \frac{K_2}{2} \sum_{i=1}^N y_i^2 - \epsilon \sum_{i=1}^N x_iy_i, \label{eqH}\end{equation}
where we take $K_1K_2>\eps^2$, otherwise the system has no bounded
ground state. We define the following extensive quantities:

\be E_1 = \frac{K_1}{2} \sum_{i=1}^N x_i^2~,~~~~~E_2 = \frac{K_2}{2}
\sum_{i=1}^N y_i^2~,~~~~~Q = \sum_{i=1}^N x_i y_i \label{q}
\ee where $ E_1 $ and $ E_2 $ are the energy of the bare systems while
$Q$ is the overlap between them.  In this case we also consider
Monte-Carlo dynamics, using the Metropolis algorithm for the
transition probability. Although the random changes in the degrees of freedom
$x_i,y_i$ are defined in the same way we have explained in the
previous section for the case of a single system, there are
different ways to implement the dynamics depending on the
updating procedure of the variables $x_i,y_i$. In this letter we will analyze
two different types of dynamics:

\begin{enumerate}
\item{\em Uncoupled or sequential dynamics.} In this case the two sets of variables $x$
and $y$ are sequentially updated. First the $x_i$ variables are updated
and the move is accepted according to the total change of energy $\Delta
E= \Delta E_1-\eps\Delta Q$. Next, the variables $y_i$ are changed and
the move accepted according to the energy change $\Delta E= \Delta
E_2-\eps\Delta Q$. This procedure is then iterated. In this
case, the dynamics does not affect simultaneously the two sets of
variables but each set is updated independently from the other. The only
coupling between the two sets of oscillators comes from the explicit
coupling term $\eps Q$ in the Hamiltonian. For $\eps=0$ the
dynamics is trivial because the dynamical evolutions are that of two
independent sets of harmonic oscillators everything reducing to the
original model described in section II.

\item{\em Coupled or parallel dynamics.} In this case the $x_i,y_i$
variables are updated in parallel according to the rule $x_i\to
x_i+r_i/\sqrt{N}$, $y_i\to y_i+s_i/\sqrt{N}$ where $r_i$ and $s_i$ are
random variables Gaussian distributed with zero average and variance
$\Delta_1$ and $\Delta_2$ respectively. The transition probability for
that move $W(\Delta E)$ is determined by the change in the total
energy $\Delta E=\Delta E_1+\Delta E_2-\eps \Delta Q$ introducing, on
top of the explicit coupling term $\eps Q$ in the Hamiltonian, an
additional coupling between the whole set of oscillators through the
parallel updating dynamics. Contrarily to the uncoupled case, the
$\eps=0$ case is interesting by itself because it shows how this kind
of dynamical coupling strongly influences the glassy behavior. In
fact, in the limiting case $\eps=0$, there will be some changes which
make the energy of one of the two systems increase, this change being
accepted because the total energy will decrease. Because of that,
despite of the fact that there is no direct coupling in the
Hamiltonian the dynamics turns out to be strongly coupled.

\end{enumerate}

\noindent

Our interest will focus on the behavior of two-times quantities such
as correlations, responses and the corresponding effective
temperatures. These quantities will refer to three classes of systems:
the set of oscillators described by the $x$ variables, the set of
oscillators described by the $y$ variables and the whole set of $x$
and $y$ variables. In the rest of the paper the subindex 1
will refer to quantities describing the set $x$ of oscillators, the
subindex 2 will refer to quantities describing the set $y$ of
oscillators and the subindex $T$ will refer to quantities describing
the whole set of oscillators $x$ plus $y$. The main set of correlation
and response functions we are interested in are:

\begin{itemize}
\item{\em Correlations.} The correlation function for the sets $x$ and
$y$,

\be 
C_1(t,s) =\frac{1}{N} \sum_{i=1}^N x_i(t)
 x_i(s)~~,~~~C_2(t,s) =\frac{1}{N} \sum_{i=1}^N y_i(t) y_i(s)~,
\label{a}
\ee                     
as well as the global correlation $C_T(t,s)=\frac{1}{2}(C_1(t,s)+C_2(t,s))$. 

\item{\em Response functions.}
The response function for the sets $x$ and $y$ are defined in the
following way. Consider two external fields $h_1$ and $h_2$ conjugated
respectively to the magnetizations $M_1=\frac{1}{N}\sum_{i=1}^N x_i $ and $M_2=\frac{1}{N}\sum_{i=1}^N y_i $, then:

\begin{equation}
 H = \frac{K_{1}}{2} \sum_{i=1}^N x_i^2 + \frac{K_2}{2} \sum_{i=1}^N y_i^2 -
 \sum_i(h_1x_i + h_2y_i) - \eps \sum_i x_iy_i~.
\end{equation} 

The responses $G_1(t,s),G_2(t,s)$ measure the change in the
magnetizations $M_1(t)$ and $M_2(t)$ induced by their respective
conjugated fields $h_1$ and $h_2$ applied at time $s$. These are
defined by

\begin{equation}
G_{1,2}(t,s) = \left( \frac{\delta M_{1,2}(t)}{\delta h_{1,2}(s)} \right)_{h_{1,2}=0}\label{eqGi}~~~~~.
\end{equation} 
Apart from these two response functions we may define the global
response function $G_T(t,s)$ as the change in the global magnetization
$M_T=\frac{1}{2}(M_1+M_2)$ induced by a field conjugate to the total magnetization,

\begin{equation}
G_T(t,s) = \left( \frac{\delta M_T(t)}{\delta h(s)}
\right)_{h=0}=\frac{1}{2}\Bigl( G_1(t,s)+G_2(t,s)\Bigr)
\end{equation} 
 
\item{\em Effective temperatures.} From the correlation and response
functions we may define three effective temperatures:  $T_{\rm eff}^1$ for the system
1, $T_{\rm eff}^2$ for system 2 and $T_{\rm eff}^T$ for the
global system. These are defined as follows,

\be T_{\rm eff}^1 = \left( \frac{\frac{\partial C_1(t,s)}{\partial
s}}{G_1 (t,s)} \right),~T_{\rm eff}^2 = \left(
\frac{\frac{\partial C_2(t,s)}{\partial s}}{G_2 (t,s)}
\right),~T_{\rm eff}^T = \left( \frac{\frac{\partial
C_T(t,s)}{\partial s}}{G_T (t,s)} \right).
\label{T1}
\ee

We will analyze in detail the three effective temperatures for the
coupled and the uncoupled cases, and show in which cases, depending on
the dynamics, these effective temperatures equalize.

\end{itemize}

\section{RESULTS}

In this section we will summarize and analyze the results obtained in Ref.\cite{art1} where the interested reader will find all the technical details.
It can be shown that the equilibrium results are the expected ones. Independent of the dynamics, the effective temperatures are just the temperature of the bath:
\be
T_{\rm eff}^1 = (2E_1 - \epsilon Q)/N = T,~~~
T_{\rm eff}^2 = (2E_2 - \epsilon Q)/N = T.
\ee
Where, in equilibrium, the energies of the subsystems are:
\be E_1^{\rm eq}=E_2^{\rm eq}=
\frac{K_1K_2T}{2(K_1K_2-\epsilon^2)}N~,~~~~~ Q^{\rm
eq}= \frac{2\epsilon E_1^{\rm
eq}}{K_1K_2}N~~.\label{eqQ} \ee
Now we analyze the results for the effective temperatures for the two
different dynamics defined in the off-equilibrium regime.

\subsection{Sequential case} 

The first quantity we have to focus on is the energy of each of the
two subsystems. In this case we can find an asymptotic solution for
the dynamical equations \cite{art1}. At first
order in logarithmic corrections $1/\log(t)$ we find in the limit $ \epsilon
\approx 0: $

\be
E_1=\frac{K_1^2K_2\Delta_1^2}{16(K_1K_2-\eps^2)\log(t)}N~,~~~E_2=\frac{K_1K_2^2\Delta_2^2}{16(K_1K_2-\eps^2)\log(t)}N~.  \label{energias}\ee

Solving the dynamic equations for the correlations and responses one can find an analytic expressions for the effective temperatures. Considering the two times $t,s$ both large but $t-s\ll s$. For a weak coupling (i.e $ \epsilon
\approx 0) $ the value of the effective temperatures are, in the limit $
s \rightarrow \infty $:

\be
T_{\rm eff}^1\approx (2E_1(s))/N + {\mathcal{O}}(\eps^2) \approx \frac{K_1^2K_2\Delta_1^2}{8(K_1K_2-\eps^2)\log(t)} \label{D11}
\ee
\be
T_{\rm eff}^2\approx (2E_2(s))/N + {\mathcal{O}}(\eps^2) \approx \frac{K_1K_2^2\Delta_2^2}{8(K_1K_2-\eps^2)\log(t)}~~~~~.\label{D12}
\ee

This yields in the $s\to\infty$ limit a non vanishing relative
difference $T_{\rm eff}^1/T_{\rm eff}^2-1$. This is a consequence of
the fact that the two energies are different in the long-time regime
(\ref{energias}). For each subsystem the effective temperature
verifies the equipartition theorem for long times. This means that, at
any time, each subsystem can be considered as if it were at
``quasi-equilibrium'' at their corresponding effective temperature.
The fact that the two effective temperatures are different implies
that there are some degrees of freedom hotter than others. One can
then imagine that there is always some kind of heat transfer or
current flow going from the ``hot degrees'' of freedom to the ``cold''
ones. Then, one may ask why the effective temperatures do not
asymptotically equalize. The reason is that the off-equilibrium
conductivity may vanish with time fast enough for the heat transfer
not to be able to compensate such difference \footnote{A detailed
study can be found in \cite{art2}}. In this situation, if we now
compute the total effective temperature for the whole system we see
that in the off-equilibrium regime this temperature does not coincide
with the sum of the energies of the systems. This is the same
situation we found in equilibrium. If we have two systems in local
equilibrium at different temperatures and put them in contact, the
global system never verifies FDT unless the two temperatures are the
same. In our case, we have two systems which are in
``quasi-equilibrium'' at two different effective temperatures, so the
$ T_{\rm eff}^T $ would never be the sum of the two energies unless
the two effective temperatures $T_{\rm eff}^1,T_{\rm eff}^2$ were the
same. In other words, two systems thermodynamically stable at
different temperatures are not globally stable when put in contact.

\subsection{Parallel dynamics}

As in the case without coupling, the interesting dynamics is when the
temperature of the bath is zero. In this case, the energies and the
overlap decay to zero logarithmically:
\beq
E_1=\frac{K_1K_2J}{8(K_1K_2-\eps^2)\log(t)}N~~~~~,~~~~~E_2=\frac{K_1K_2J}{8(K_1K_2-\eps^2)\log(t)}N~,\label{eqIVaa} \nonumber\\
Q=\frac{\eps J}{4(K_1K_2-\eps^2)\log(t)}N~~{\rm with}~~J=\frac{K_1\Delta_1^2}{2}+\frac{K_2\Delta_2^2}{2}  . 
\eeq
For the effective temperature up to order $\eps^2 $ we may write, in the limit $ s \longrightarrow
\infty $ (with $\frac{t}{s}$ finite):

\be T_{\rm eff}^1=(2E_1(s) - \eps Q(s))/N~,~~~~~T_{\rm eff}^2=(2E_2(s) -
\eps Q(s))/N, \ee because the asymptotic values of the $ E_1(s) $ and $
E_2(s) $ are the same the  effective temperatures
for the subsystems are also the same in the long-time limit. Note that the case with dynamic coupling or
parallel dynamics is qualitatively different from the case without
dynamic coupling or sequential, because now all the degrees
of freedom are at the same effective temperature in the long-time
limit. As a result, if we consider the global system, the total effective temperature defined in (\ref{T1}) is, in the
limit $ s\rightarrow \infty $ with $ \frac{t}{s}$ finite:

\be T_{\rm eff}^T=T_{\rm eff}^1=T_{\rm eff}^2=(2E'(s) - \eps Q(s))/N \ee
where $ E'=E_1=E_2 $ and $Q$ are given by (\ref{eqIVaa}). This is a
consequence of the fact that the energies of the two systems equalize
due to the dynamic coupling. Now the dynamic coupling equalize the energies of the subsystems and the whole system has the same
effective temperature and we can define an effective temperature for
the global system using FDT. The situation is the same as in
equilibrium systems. If we have two systems in equilibrium at a
certain temperature $T$, FDT not only holds for each subsystem but
also holds for the whole system bringing the temperature of the bath
$T$. 
If we restrict to the case in
which the coupling constant vanishes, $\eps=0$, then the systems are still
coupled only through the dynamics and we obtain the same qualitatively
results: $ T_{\rm eff}^T=T_{\rm eff}^1=T_{\rm eff}^2=(2E'(s))/N $  with $
E'(s)\approx \frac{J}{8\log(s)}N$. We conclude that the dynamic
coupling does not allow the presence of more than one effective
temperature in the whole system because even in the absence of explicit
coupling in the Hamiltonian, the dynamics itself makes the energies to
equalize in the long-time limit regime.

\section{CONCLUSIONS}

In this paper we have focused our attention on the concept of the
effective temperature defined through the FDR (\ref{eqX}). The
effective temperature, a parameter defined as an extension of FDT to
the off-equilibrium regime has been introduced in the context of glass
theory in order to understand the physics behind the dynamic behavior
of these out-off-equilibrium systems. In this paper we hope to have
clarified some aspects behind the physical meaning of this effective
temperature. We have studied two types of couplings between the two
subsystems of oscillators, both in an aging state, finding that the
way we couple them is crucial for the validity of the zeroth law in
the off-equilibrium regime to hold.  The two cases we studied are the
dynamically uncoupled or sequential case and the dynamically coupled
or parallel case. In short, for the sequential case the coupling
between the variables of the two subsystems in the resulting dynamics
arises only through the Hamiltonian term $\eps Q$. For the parallel
case, the variables of the two subsystems are simultaneously updated
leading to further interaction between the two subsystems (on top of
the $\eps Q$ coupling term in the energy).

For the dynamically uncoupled or sequential case the two subsystems
asymptotically reach different effective temperatures which never
equalize. So we can divide the system into two parts, each part
characterized by its own effective temperature. The reason for this
behavior is that off-equilibrium thermal conductivity decays very
quickly to allow for an asymptotic equalization of the two effective
temperatures. In fact, it decays as \cite{art2}: \be J\approx
\frac{1}{t\log^2(t)}\left( \frac{1}{T_{\rm eff}^1}-\frac{1}{T_{\rm
eff}^2}\right). \ee 

Physically this means that, as time goes on, although the effective
temperatures of the systems are different the conductivity is not high
enough and they cannot thermalize. This seems to be the reason for the
existence of different temperatures in real glasses, in which there
are {\it fast} degrees of freedom and {\it slow} ones at different
effective temperatures which never equalize due to the smallness of
the conductivity. Our conclusion is that the zeroth law is probably
valid but hardly effective due to the very small conductivity between
the two subsystems in the aging state.

We have seen that for the dynamically coupled or parallel case, the
two effective temperatures equalize and the two subsystems are in a
sort of thermal equilibrium between them in the aging state. As a
result, all degrees of freedom have the same temperature. In this
case, the direct coupling of the two subsystems through the parallel
dynamics makes the conductivity much larger than in the sequential
case so in this case a zero-th law for the aging state is effective
and holds. In fact, these results are also valid when we consider the
particular case $ \eps = 0 $ in which the dynamics in itself is enough
to equalize the effective temperatures.

Dynamics in real structural glasses involves short scale motions of
atoms and coupling between the different degrees of freedom occurs at
the level of the energy or Hamiltonian and never at the level of the
dynamics. Then, from these two types of couplings the first one is the only realistic. 

The results of this paper explain then why different degrees of
freedom in structural glasses can stay at different effective
temperatures forever. The off-equilibrium conductivity or heat
transfer between the different degrees of freedom is small enough for
the equalization of the effective temperatures associated to the
different degrees to never occur. This explains why when we touch a
piece of glass we feel it at the room temperature. To conclude,
although a zero-th law for non-equilibrium glassy systems may hold, it
is hardly effective because of the small energy transfer occurring
between degrees of freedom at different effective temperatures.  It
would be very interesting to pursue this investigation further by
studying other solvable examples and showing that what we have
exemplified here is a generally valid for structural glasses as well
as for other glassy systems.

{\bf Acknowledgements.}  It is a pleasure to thank F. Ritort for
proposing this work and for very interesting discussions. This work is
supported by a grant from the University of Barcelona.


\section*{References}

\begin{thebibliography}{00}


\bibitem{SIT} Proceedings of the XIV Sitges Conference, ``Complex Behavior of Glassy Systems'', M.Rubi and C.Perez-Vicente Eds., (Springer-Verlag, Berlin, 1997)

\bibitem{BOU} J-P Bouchaud,L.F.Cugliandolo, J.Kurchan and M.M\'ezard, ``Out of equilibrium dynamics in spin-glasses and other glassy systems, Preprint cond-mat/ \textbf{9702070}, in 'Spin-glasses and random fields, A.P.Young ed. (World Scientific, Singapore).

\bibitem{KUB} R.Kubo, Rep.Progr.Phys \textbf{29} 255 (1966); R.Kubo, M.Toda and N.Hshitsume, Statistical Physics II (2nd ed.) Springer Verlag, Berlin (1991).

\bibitem{CUG} L. F. Cugliandolo and J. Kurchan, Phys. Rev. Lett.\textbf{71}, 173
(1993); J.Phys. A (Math. Gen.) {\bf 27} 5749 (1994).


\bibitem{FRA} S. Franz and M. M\'ezard, Europhys. Lett. {\bf 26}, 209 (1994);
Physica A {\bf 210}, 48 (1994).

\bibitem{FMPP} S. Franz, M. Mezard, G. Parisi and L. Peliti,
Phys. Rev. Lett. {\bf 81} (1998) 1758; J. Stat. Phys. {\bf 97}
(1999)459




\bibitem{KUR} L.F.Cugliandolo, J.Kurchan and L.Peliti,
Phys. Rev. E,{\bf 55}, 3898 (1997).

\bibitem{CUKU} L. F. Cugliandolo and J. Kurchan, Physica {\bf A263}, 242 (1999); Preprint cond-mat/{\bf 9911086}. 

\bibitem{EXH} R.Exartier and L.Peliti, Preprint cond-mat/ \textbf{9910412}.



\bibitem{NIEW2} Th.M.Nieuwenhuizen, Phys. Rev.Lett.\textbf{80}, 5580 (1998).  

\bibitem{EXP} T.S.Grigera and N.E.Israeloff, Phys. Rev. Lett. {\bf 83},
5038 (1999); L. Bellon, S. Ciliberto and C. Laroche, {\em
Fluctuation-Dissipation-Theorem violation during the formation of a
colloidal-glass} Preprint cond-mat/\textbf{0008160}

\bibitem{art2} A.Garriga and F.Ritort, Eur. Phys. J. B, {\bf 21}, 115, (2001). 

\bibitem{BPR} L.L.Bonilla, F.G. Padilla and F.Ritort, Physica A,\textbf{250},315 (1998)


\bibitem{art1} A.Garriga and F.Ritort, Eur. Phys. J. B, {\bf 20}, 105-122, (2001). 



\bibitem{TEO} Th.M.Nieuwenhuizen, Phys. Rev. E {\bf 61}, 267 (2000)
\bibitem{FEL} A. Crisanti and F. Ritort,  Preprint cond-mat/ \textbf{0009261}.










\end{thebibliography}
\end{document}